\documentclass[10pt,twocolumn,letterpaper]{article}

\usepackage{wacv}              

\definecolor{wacvblue}{rgb}{0.21,0.49,0.74}
\usepackage[pagebackref,breaklinks,colorlinks,allcolors=wacvblue]{hyperref}

\def\wacvPaperID{457} 
\def\confName{WACV}
\def\confYear{2027}

\title{NAIMA: Semantics Aware RGB Guided Depth Super-Resolution}

\author{Tayyab Nasir\\
The University of Western Australia\\
Perth, Western Australia\\
{\tt\small tayyabnasir22@gmail.com, tayyab.nasir@research.uwa.edu.au}
\and
Daochang Liu\\
The University of Western Australia\\
Perth, Western Australia\\
{\tt\small daochang.liu@uwa.edu.au}
\and
Ajmal Mian\\
The University of Western Australia\\
Perth, Western Australia\\
{\tt\small ajmal.mian@uwa.edu.au}
}

\begin{document}

\maketitle

\begin{figure*}[!htb]
  \includegraphics[width=\textwidth]{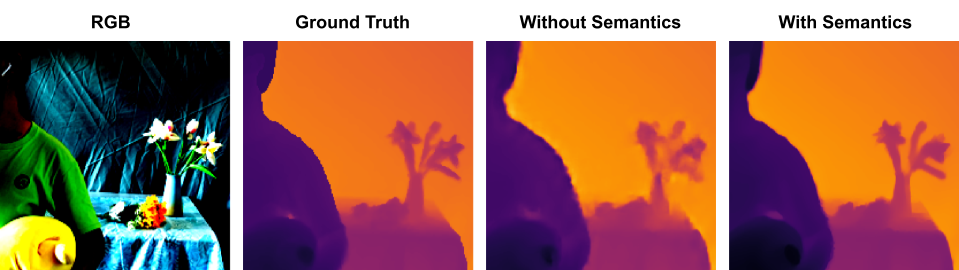}
  \caption{Blurred depth discontinuities caused by RGB noise when performing super-resolution without semantic guidance. In contrast, our semantics-aware approach leverages global contextual information to better preserve structural fidelity.}
  \label{fig:teaser}
\end{figure*}
\begin{abstract}
Guided depth super-resolution (GDSR) is a multi-modal approach for depth map super-resolution that relies on a low-resolution depth map and a high-resolution RGB image to restore finer structural details. However, the misleading color and texture cues indicating depth discontinuities in RGB images often lead to artifacts and blurred depth boundaries in the generated depth map. Recent methods counter this by drawing priors from large pretrained models, but these priors enter the network as decoded predictions such as relative depth, surface normal, or segmentation maps, coupling restoration quality to the accuracy of the decoded prior and requiring auxiliary objectives. We propose a solution that introduces global contextual semantic priors, generated from pretrained vision transformer token embeddings, injecting them directly into the depth branch. Our Guided Token Attention (GTA) module lets multi-level depth encodings act as queries of cross-attention over semantic tokens drawn from progressively deeper layers of the pretrained visual transformer (ViT), scaled by a zero-initialized gate that admits semantic evidence only to the extent that it reduces reconstruction error. The resulting token-to-depth correspondence is aligned implicitly, through learned attention under a single reconstruction loss, rather than through an explicit alignment or distribution-matching objective. Building on this, we present Neural Attention with Implicit Multi-token Alignment (NAIMA), which, to the best of our knowledge, is the first GDSR framework guided by undecoded semantic tokens. NAIMA remains competitive in-distribution while achieving the strongest cross-dataset generalization\footnote{Code available at: \href{https://github.com/tayyabnasir22/NAIMA-GDSR}{ https://github.com/tayyabnasir22/NAIMA-GDSR}}.
\end{abstract}

\section{Introduction}
Depth maps are crucial for numerous computer vision tasks, including 3D scene reconstruction, object detection, and semantic segmentation~\cite{wang2024sgnet, yuan2023structure, zhong2023guided}. They are widely utilized in real-world applications such as autonomous driving and industrial manufacturing systems~\cite{zhong2023guided, yuan2023recurrent, wang2025dornet}. Compared to RGB sensors, depth sensors still lag in terms of resolution, accuracy, and cost-effectiveness. Hence, RGB-guided depth map super-resolution (GDSR) has emerged as a practical solution to enhance depth resolution without the need for expensive hardware.

GDSR is motivated by the observation that RGB images and depth maps provide complementary geometric representations of the same scene~\cite{zhong2023guided, cong2018hscs, cong2016saliency}. Under this assumption, the objective is to integrate low-frequency structural information from the blurred low-resolution depth map with high-frequency features of the corresponding high-resolution RGB image to reconstruct a high-resolution depth map. However, an RGB image contains two distinct kinds of edges: geometric edges, which correspond to true depth discontinuities between surfaces, and photometric edges, which arise from texture, color changes, and shadows, which do not represent depth discontinuities. Distinguishing the two requires reasoning about object and region identity rather than local appearance. Several studies have highlighted this cross-modal misalignment as a fundamental challenge of the multi-modal setting~\cite{wang2024sgnet, tang2021bridgenet, li2020asif, li2020rgb}, where in the absence of higher-level guidance, a naive fusion of spatial RGB features with LR depth features leads to texture copying, blurred object boundaries, and misaligned geometric structures. Usually, the photometric edges occur within a single object or surface, whereas true depth discontinuities occur at boundaries between distinct objects. Hence, incorporating semantic information alongside RGB spatial features is crucial for disambiguating photometric edges from genuine depth discontinuities.

A large number of recent works have augmented GDSR with knowledge beyond RGB-depth features alone, such as multi-task learning, which jointly optimizes GDSR with auxiliary objectives such as monocular depth estimation, gradient prediction, or self-supervised degradation modeling~\cite{tang2021bridgenet, wang2024sgnet, wang2025dornet}. Recent works have also leveraged large foundation models as prior generators for GDSR~\cite{yan2025ducos, wang2026scene}. However, these methods first require semantic or geometric priors to be decoded into task-specific predictions, such as relative depths~\cite{yan2025ducos}, surface normals, or segmentation maps~\cite{wang2026scene}, before this additional knowledge from the priors can be incorporated into the reconstruction process. This couples restoration quality to the accuracy of the decoded prediction, and demands additional optimization machinery, such as constraint terms~\cite{yan2025ducos} or distillation losses~\cite{wang2026scene}, to keep the decoded signal from destabilizing reconstruction. Depth Anything~\cite{yang2024depth} exploited such semantics without decoding, aligning its encoder features with frozen DINOv2~\cite{oquab2023dinov2} embeddings. However, it operates in a single-modality estimation setting, where semantically enriched RGB features are decoded directly into relative depth.

\begin{figure*}[!htb]
  \centering
  \includegraphics[width=0.9\linewidth]{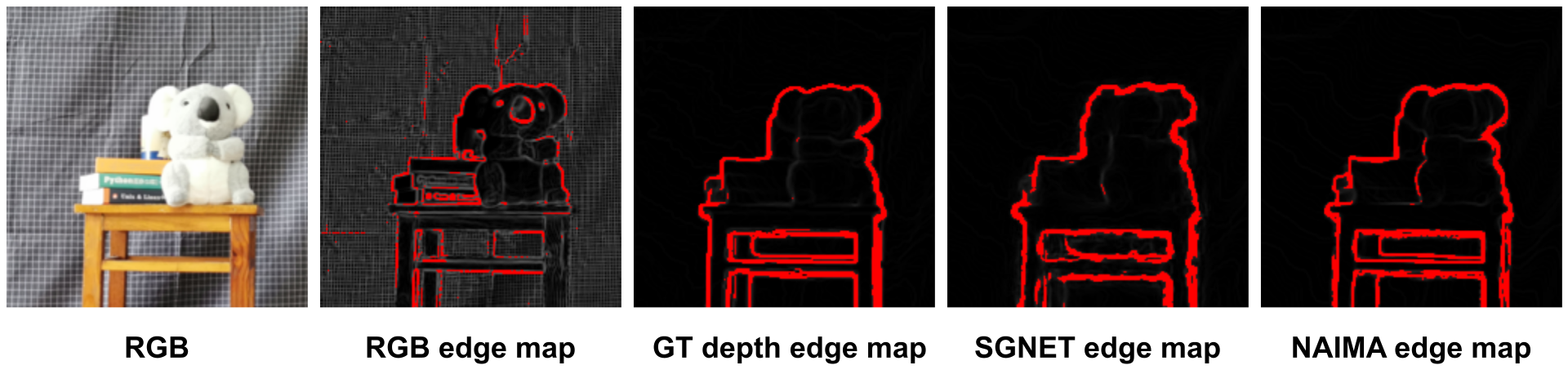}
  \caption{Edge maps overlaid on the RGB image, the ground-truth depth map, and depth maps predicted by different super-resolution techniques. The semantically guided prediction preserves sharper boundaries and avoids the boundary-bleeding problem, while leveraging object semantics to recover true geometric boundaries, even where the low-resolution input is missing information.}
    \label{fig:edges}
\end{figure*}

We propose NAIMA (Neural Attention with Implicit Multi-token Alignment), which, for the first time, uses raw, undecoded semantic tokens themselves in a multi-modal GDSR problem, refining the depth geometry already measured in the low-resolution depth input rather than generating one from RGB alone. NAIMA operates in multiple stages, drawing semantic tokens of different abstraction levels from progressively deeper layers of the pretrained ViT. The alignment between tokens and depth is learned implicitly by the proposed Guided Token Attention (GTA) module, which lets encoded depth features act as queries of a cross-attention operation over projected semantic tokens, so the depth stream actively selects semantic evidence for missing structure without overwriting the geometry it already holds. A zero-initialized learnable gate scales the attended residual so that each stage starts as an identity mapping and admits semantics only to the extent that it reduces reconstruction error. The semantically refined depth encodings are then fused with complementary spatial RGB features, and the entire framework is trained with a single reconstruction objective, without the need for an alignment loss, constraint optimization, or distillation term. Figure~\ref{fig:edges} illustrates how this semantic guidance recovers true geometric boundaries and avoids boundary bleeding, even where the low-resolution input is missing information.

Our contributions are summarized as follows:
\begin{itemize}
    \item We introduce NAIMA, to the best of our knowledge the first GDSR framework to exploit raw, undecoded semantic tokens of a frozen self-supervised vision transformer as guidance. Unlike recent foundational model-based methods that decode priors into intermediate predictions~\cite{yan2025ducos, wang2026scene}, our design decouples restoration quality from the accuracy of any decoded prediction.
    \item We propose the Guided Token Attention (GTA) module, in which multi-level depth encodings act as queries of a cross-attention over projected semantic tokens, gated by a zero-initialized residual scale. GTA aligns depth features to true object boundaries implicitly, allowing the entire framework to be trained with a single reconstruction objective.
    \item Extensive experiments on the Middlebury~\cite{scharstein2003high}, LU~\cite{lu2014depth}, NYU\_v2~\cite{silberman2012indoor}, and RGBDD~\cite{he2021towards} datasets against 11 benchmark methods demonstrate that NAIMA offers strong cross-dataset generalization, achieving the lowest RMSE in six of nine settings and ranking first or second in 11 of 12 settings overall, while remaining competitive in-distribution at a fraction of the guidance-model cost of prior foundation-model approaches.
\end{itemize}

\section{Related Work}
Obtaining high-quality depth data is costly and often constrained by hardware limitations~\cite{ariav2022depth}. To address this limitation, prior work has largely focused on multi-modal approaches, where a high-resolution RGB image serves as a guiding modality to recover missing high-frequency details. In recent years, various techniques have been proposed for both standard and RGB-guided depth super-resolution, introducing end-to-end deep learning architectures to reconstruct high-resolution depth maps from low-resolution inputs. Early works such as~\cite{zuo2021mig, kim2021deformable, hui2016depth, de2022learning} were among the first to employ deep neural networks for guided depth super-resolution. These methods typically apply convolutional neural networks to fused features extracted from the low-resolution depth map and the high-resolution RGB guide to reconstruct a single high-resolution depth output.
However, a key challenge in RGB-guided depth super-resolution remains the presence of noise in RGB images, such as misleading color and texture patterns that do not necessarily correspond to actual geometric shapes. This, in turn, can cause blurred boundaries and artifacts in the resultant high-resolution depth map. This issue has been widely recognized in prior work, with many novel architectures proposing mechanisms such as attention-based fusion, auxiliary learning, and constrained feature integration to mitigate noise from RGB guidance~\cite{li2020asif, wang2024sgnet, wang2025dornet, tang2021bridgenet}. These approaches aim to selectively filter irrelevant RGB information and focus on strongly correlated features to improve reconstruction quality. Prior works have used attention for learned feature fusion mechanisms to better model the correlation between RGB and depth modalities~\cite{zhong2021high, yang2022codon, song2020channel, yan2025ducos, wang2026scene}. By leveraging attention, these approaches aim to suppress irrelevant texture artifacts, selecting only relevant features, thereby reducing blurred boundaries and reconstruction errors caused by noisy RGB guidance. 
In addition, many studies adopt multi-task learning strategies to further regulate the interaction between RGB and depth features. For example, monocular depth estimation (MDE) has been used as an auxiliary task to enhance shared feature representations, improving overall GDSR performance~\cite{tang2021bridgenet}. Methods such as SGNet~\cite{wang2024sgnet} and DORNet~\cite{wang2025dornet} introduce gradient–frequency estimation and self-supervised degradation modeling, respectively, improving robustness under real-world degraded depth conditions. Similarly, another such study~\cite{yan2022learning} incorporates depth completion as an auxiliary task, arguing that such supervision helps capture contextual dependencies and better control feature interactions between RGB and depth modalities. While effective, these mechanisms all derive their guidance from the RGB-depth pair itself, and the auxiliary objectives add optimization complexity to the training pipeline.
The challenge of noisy texture and color information in RGB has also been widely studied in monocular depth estimation (MDE), where semantic segmentation is commonly adopted as an auxiliary task to enforce global semantic consistency~\cite{chen2019towards, guizilini2020semantically}. Recent architectures such as Depth Anything~\cite{yang2024depth} extend this idea by aligning encoder features with frozen DINOv2~\cite{oquab2023dinov2} embeddings, demonstrating that pretrained token representations can replace explicit semantic supervision. Within GDSR, foundation-model priors have only recently been explored~\cite{yan2025ducos, wang2026scene}. However, these methods rely on decoded predictions, such as relative depth, surface normal, or segmentation maps, before the prior can guide reconstruction, coupling restoration quality to the accuracy of the decoded output and requiring auxiliary objectives. The use of undecoded, token-level semantic representations as guidance in GDSR remains unexplored, which our proposed NAIMA framework addresses.

\section{Methodology}
\begin{figure*}[!htb]
  \centering
  \includegraphics[width=0.98\linewidth]{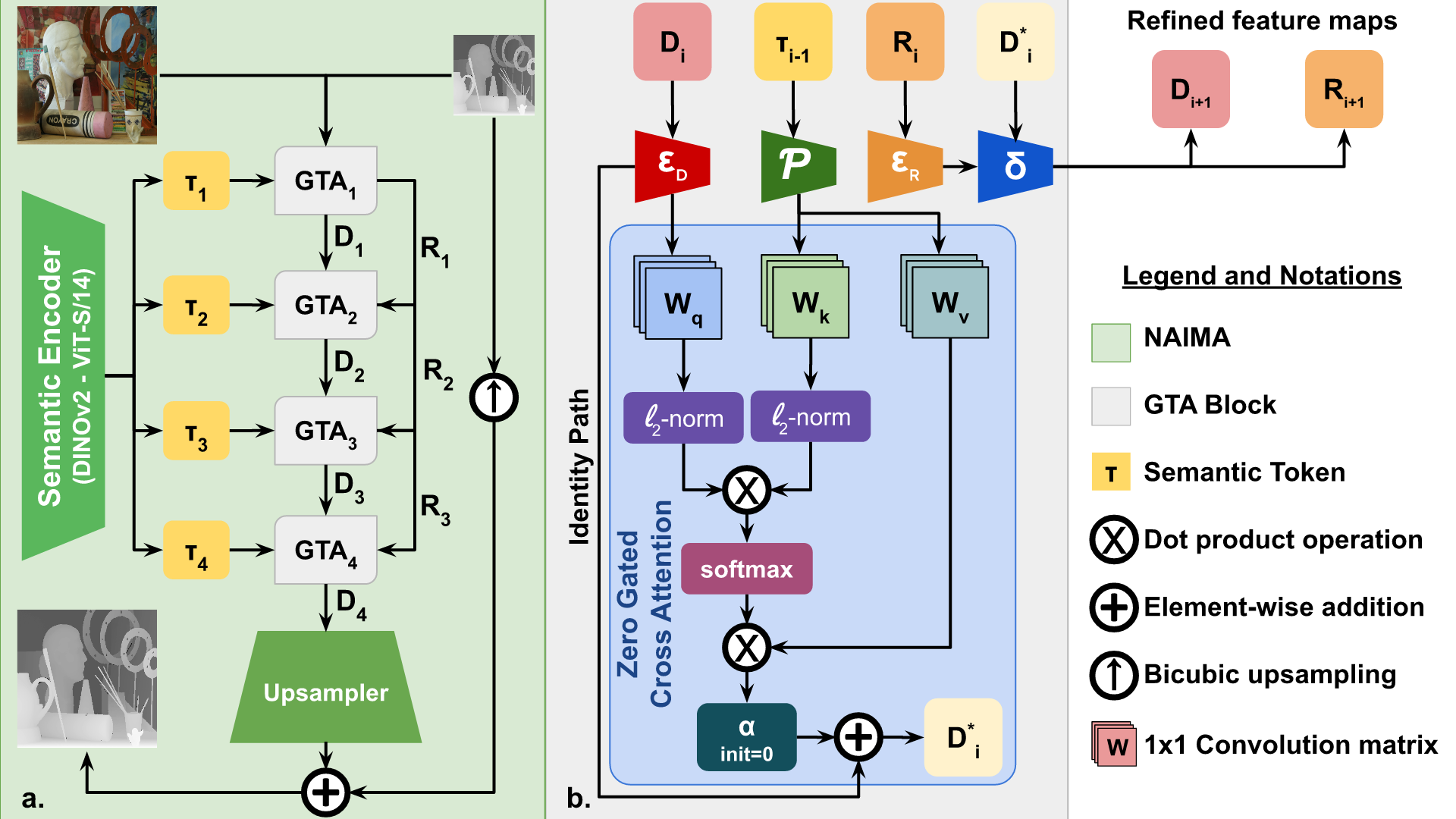}
  \caption{Overview of NAIMA. \textbf{a.} Full architecture: a frozen DINOv2-ViT-S/14 provides multi-level semantic tokens, injected into four Guided Token Attention (GTA) blocks that progressively refine depth features alongside a recurrent RGB stream. \textbf{b.} A single GTA block: depth features query L2-normalized projections of the semantic tokens via cross-attention, scaled by a zero-initialized gate $\alpha$, added back through the identity path, and fused with spatial RGB features.}
    \label{fig:archi}
\end{figure*}

Figure\hyperref[fig:archi]{~\ref*{fig:archi}a)} presents a high-level overview of the NAIMA architecture. The input to the model is a high-resolution RGB image $I \in \mathbb{R}^{3 \times H \times W}$ and a low-resolution depth map $D_{lr} \in \mathbb{R}^{1 \times h \times w}$, where $h = \frac{H}{s}$, $w = \frac{W}{s}$, and $s$ denotes the upscaling factor.

Semantic information is extracted from $I$, by passing it through a pretrained DINOv2, obtaining intermediate patch-level tokens $\left\{\tau_{1}, \tau_{2}, \tau_{3}, \tau_{4}\right\}$. Each token embedding is extracted from a different transformer block and encodes progressively higher-level semantics. We adopt DINOv2 specifically because its self-supervised patch tokens are dense, object-aware, and largely texture-invariant, needed to separate geometric edges from photometric edges. The same characteristic is beneficial for monocular depth estimation, as demonstrated by Depth Anything~\cite{yang2024depth}. To ensure that the frozen semantic encoder does not dominate the trainable components of the model, we use the lightest DINOv2 (ViT-S/14) variant, where implementation details are presented in Section~\ref{subsec:implementation}, with further discussion of the semantic encoder and extracted semantic tokens presented in the appendix.

Depth features are iteratively encoded using the depth encoder $\mathcal{E}_{D}$, built upon residual channel attention blocks~\cite{zhang2018image}, which have been widely adopted in the literature for spatial feature representation. Let $E_{i}$ denote the enriched depth feature map at level $i \in \left\{1, 2, 3, 4 \right\}$, with $D_{0} = D_{lr}$. The iterative encoding is defined as: 

{\small
\begin{equation}
    E_{i} = \mathcal{E}_D(D_{i - 1}).    
\end{equation}
}

\subsection{Guided Token Attention}
Figure\hyperref[fig:archi]{~\ref*{fig:archi}b)} presents the details of a single GTA block. The core idea behind our GTA module is to inject higher-level semantic information from the token embeddings into the depth feature maps before fusion with spatial RGB features. Semantics first establishes where true object boundaries lie, so that the subsequent RGB fusion transfers high-frequency detail only along geometry-consistent structures; reversing this order would expose the depth stream to photometric edges before it has the semantic context to reject them. For this, we employ cross-attention to compute the correlation between the encoded depth features $E_{i}$ (Query) and the RGB tokens $\tau_{i}$ (Key and Value). The semantic tokens are first reshaped into their corresponding 2D spatial grid and then undergo convolutional projections to introduce local spatial interactions:

\begin{equation}
    S_{i} = \mathcal{P}(\tau_{i}),
\end{equation}

\noindent where $S_{i}$ represents the projected semantic feature map.

The projected semantics map $S_{i}$ is then passed through a pixel shuffle operation $\mathcal{U}$, which converts channel-wise sub-pixel encodings into a dense, high-resolution spatial representation, ensuring that the semantic features have the same spatial dimensions as the encoded depth map $E_{i}$.

\begin{equation}
    F_{i} = \mathcal{U}(S_{i}),
\end{equation}

Next, the encoded depth feature map $E_{i}$ queries the projected semantic features $F_{i}$ through cross-attention. Both modalities are first mapped into a shared interaction space through learned projections. Feature maps are flattened along the spatial dimension, and the queries and keys are L2-normalized before the dot product. This bounds the attention logits and, together with the zero-initialized gate, stabilizes the cross-modal interaction. 

\begin{equation}
    \label{eq:qkv}
    Q_{i} = W_{q}E_{i}, K_{i} = W_{k}F_{i}, V_{i} = W_{v}F_{i},
\end{equation}

\noindent and the attended semantic context is injected into the depth features through a residual gated connection:

\begin{equation}
\label{eq:residual_attention}
    D^{*}_{i} = E_{i} + \alpha * Softmax\left(\frac{Q_{i}K_{i}^{T}}{\sqrt{d_{k}}}\right)V_{i},
\end{equation}

\noindent where $W_{q}$, $W_{k}$, $W_{v}$ are learned projection weights that map the heterogeneous features into a shared interaction space, and $d_{k}$ is the dimension of the key vectors. The scalar $\alpha$ is a learnable gate initialized to zero, so that the depth stream initially passes through unmodified, and semantic context is gradually introduced as training progresses to reduce the reconstruction error. This design guarantees that the frozen semantic prior cannot destabilize early training, and it is what allows semantic guidance to be integrated without any auxiliary alignment or constraint objective, in contrast to prior foundational model-based approaches~\cite{yan2025ducos, wang2026scene}. Moreover, the converged per-level values of $\alpha$ offer a direct, interpretable readout of how much semantic guidance each depth encoding level consumes.

At this stage, the depth feature map $D^{*}_{i}$ has been enriched with low-frequency details from the low-resolution depth map and semantic information from the RGB tokens. To incorporate high-frequency details from the high-resolution RGB image $I$, we pass it through a simple feature encoder composed of residual blocks:

\begin{equation}
    R^{*}_{i} = \mathcal{E}_{R}(I).
\end{equation}

These spatial RGB features  $R^{*}_{i}$ are then infused with $D^{*}_{i}$, using an existing feature fusion block~\cite{shi2022symmetric}, producing the refined depth feature map at the corresponding level:

\begin{equation}
    D_{i+1} = \delta(D^{*}_{i}, R^{*}_{i}),
\end{equation}

\noindent where $D_{i+1}$ serves as the input for the subsequent GTA block. This process is repeated iteratively across all levels. The final refined depth feature map is passed through an upsampling network $\phi$, and the bicubic upsampled input low-resolution depth map is added to it to generate the high-resolution depth map $D_{hr}$.

\begin{equation}
    D_{hr} = \phi(D_{4}) + D_{lr}^{\uparrow},
\end{equation}

\subsection{Loss Function}
We adopt a gradient-aware pixel loss that consists of a pixel-wise $\mathcal{L}_{1}$ reconstruction term between the predicted and ground-truth depth maps, along with an additional $\mathcal{L}_{1}$ loss computed between their spatial gradients~\cite{nasir2026implicit}.
\begin{equation}
    \mathcal{L} = \mathcal{L}_{1} + \lambda \mathcal{L}_{grad},
\end{equation}
where
\begin{equation}
    \mathcal{L}_{1} = \parallel \hat{D}_{sr} - D_{gt} \parallel_{1},
\end{equation}
and
\begin{equation}
    \mathcal{L}_{grad} =  \parallel \nabla_{x} \hat{D}_{sr} - \nabla_{x} D_{gt} \parallel_{1} +  \parallel \nabla_{y} \hat{D}_{sr} - \nabla_{y} D_{gt} \parallel_{1},
\end{equation}

\noindent where $\lambda$ is the gradient control term set to 0.05~\cite{nasir2026implicit}.

\section{Experiments}
\subsection{Datasets}
We follow the same training and evaluation protocols as adopted in prior work~\cite{kim2021deformable, metzger2023guided, wang2024sgnet}. Our model is trained on the NYU\_v2 benchmark using scaling factors of 4$\times$, 8$\times$, and 16$\times$, where low-resolution depth maps are synthetically generated through bicubic downsampling, following the standard train-test split. For cross-dataset validation, the same models trained exclusively on the NYU\_v2 dataset are used without any fine-tuning. Evaluation is conducted on the RGBD-D, Middlebury, and LU datasets, using root-mean-squared error (RMSE) in centimeters to measure performance. Since all models are trained solely on NYU\_v2, we report NYU\_v2 as the in-distribution setting and treat RGBDD, Middlebury, and LU as out-of-distribution evaluations.

\subsection{Implementation}
\label{subsec:implementation}
For NAIMA, the DINOv2 backbone remains frozen throughout training, preserving its pretrained semantic priors and ensuring that the semantic encoder acts only as a guiding component rather than driving the reconstruction. With 22 million frozen parameters, which constitute a minority of the total parameters and are entirely non-trainable, the semantic encoder does not dominate training. Consistent with prior work, all models are trained on bicubic downsampled inputs for each scaling factor and optimized for 200 epochs. We use the Adam optimizer with an initial learning rate of 0.0001, and employ a multi-step learning rate decay scheduler. All experiments are conducted with a batch size of 1 on an NVIDIA RTX 4090 GPU.

To ensure a fair comparison consistent with prior work~\cite{wang2024sgnet, wang2026scene, wang2025dornet}, we train only on NYU\_v2 and report baseline results from their respective original publications rather than reproducing them. For the TOFDSR~\cite{yan2024tri} comparison, we likewise evaluate only models with publicly available NYU\_v2-trained weights. In both cases, we prioritize adherence to established benchmarks and evaluation protocols over any selective retraining or tuning.

\subsection{Quantitative Results}
Table~\ref{tab:main_res} reports comparative performance across four datasets and three scaling factors. In the in-distribution setting, NAIMA is on par with the strongest existing method, SPFNet, trailing by at most 0.06cm at 16$\times$ while using a single frozen ViT-S/14 encoder in place of two large decoded-prior models (Omnidata and SAM) and requiring no auxiliary objectives. In the out-of-distribution setting, NAIMA achieves the lowest RMSE in six of the nine dataset–scale configurations and ranks first or second in eight, with the advantage concentrated at $8\times$ and $16\times$, e.g., a 7.7\% RMSE reduction over the next best method on Lu at $8\times$. This pattern supports our central claim i.e., undecoded token-level guidance transfers across depth distributions better than decoded priors, whose task heads are themselves susceptible to distribution shift, and matters most precisely when the LR depth carries the least structure.

\begin{figure}[!htb]
  \centering
  \includegraphics[width=0.7\linewidth]{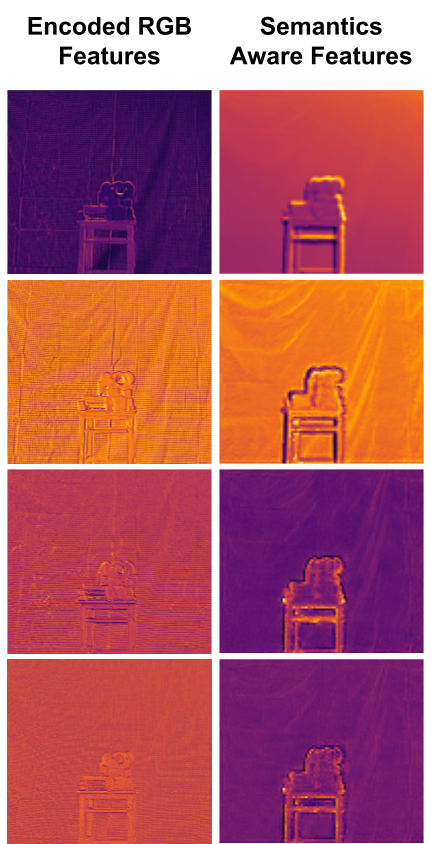}
  \caption{Visualization of feature maps extracted from different layers of the NAIMA architecture, for the model trained for an 8$\times$ scaling factor. The top row shows RGB features encoded at each level, while the bottom row shows the corresponding refined feature maps obtained by fusing RGB features with semantically guided depth representations. Feature levels increase from left to right. The semantically corrected features have less noise compared to the features coming directly from RGB spatial encoding.}
    \label{fig:feature}
\end{figure}

\begin{table*}[!htb]
    \centering
    \begin{tabular}{lccc|ccccccccc}
    \toprule
    \textbf{Methods} 
    & \multicolumn{3}{c|}{\textbf{In-Distribution}} 
    & \multicolumn{9}{c}{\textbf{Out-of-Distribution}} 
    \\
    & \multicolumn{3}{c|}{\textbf{NYU\_v2}} 
    & \multicolumn{3}{c}{\textbf{RGBDD}} 
    & \multicolumn{3}{c}{\textbf{Middlebury}}
    & \multicolumn{3}{c}{\textbf{LU}} \\
    \cmidrule(lr){2-4} \cmidrule(lr){5-7} \cmidrule(lr){8-10} \cmidrule(lr){11-13}
    & 4$\times$ & 8$\times$ & 16$\times$ & 4$\times$ & 8$\times$ & 16$\times$ & 4$\times$ & 8$\times$ & 16$\times$ & 4$\times$ & 8$\times$ & 16$\times$ \\
    \midrule
    DKN 2021~\cite{kim2021deformable} & 1.62 & 3.26 & 6.51 & 1.30 & 1.96 & 3.42 & 1.23 & 2.12 & 4.24 & 0.96 & 2.16 & 5.11 \\
    FDKN 2021~\cite{kim2021deformable} & 1.86 & 3.58 & 6.96 & 1.18 & 1.91 & 3.41 & 1.08 & 2.17 & 4.50 & \textcolor{red}{0.82} & 2.10 & 5.05 \\
    FDSR 2021~\cite{he2021towards} & 1.61 & 3.18 & 5.84 & 1.18 & 1.74 & 3.05 & 1.13 & 2.08 & 4.39 & 1.29 & 2.19 & 5.00 \\
    JIIF 2021~\cite{tang2021joint} & 1.37 & 2.76 & 5.27 & 1.17 & 1.79 & 2.87 & 1.09 & 1.82 & 3.31 & 0.85 & 1.73 & 4.16 \\
    DCTNet 2022~\cite{zhao2022discrete} & 1.59 & 3.16 & 5.84 & \textcolor{blue}{1.08} & 1.74 & 3.05 & 1.10 & 2.05 & 4.19 & 0.88 & 1.85 & 4.39 \\
    SUFT 2022~\cite{shi2022symmetric} & 1.12 & 2.51 & 4.86 & 1.10 & 1.69 & 2.71 & 1.07 & 1.75 & 3.18 & 1.10 & 1.74 & 3.92 \\
    DAGF 2023~\cite{zhong2023deep} & 1.36 & 2.87 & 6.06 & 1.18 & 1.82 & 2.91 & 1.15 & 1.80 & 3.70 & \textcolor{blue}{0.83} & 1.93 & 4.80 \\
    DADA 2023~\cite{metzger2023guided} & 1.54 & 2.74 & 4.80 & 1.20 & 1.83 & 2.80 & 1.20 & 2.03 & 4.18 & 0.96 & 1.87 & 4.01 \\
    SGNet 2024~\cite{wang2024sgnet} & 1.10 & 2.44 & 4.77 & 1.10 & \textcolor{blue}{1.64} & 2.55 & 1.15 & 1.64 & 2.95 & 1.03 & 1.61 & 3.55 \\
    DORNet 2025~\cite{wang2025dornet} & 1.19 & 2.70 & 5.60 & - & - & - & - & - & - & - & - & - \\
    SPFNet 2026~\cite{wang2026scene} & \textcolor{red}{1.09} & \textcolor{red}{2.36} & \textcolor{red}{4.55} & 1.13 & 1.71 & \textcolor{blue}{2.53} & \textcolor{blue}{1.05} & \textcolor{red}{1.57} & \textcolor{blue}{2.79} & 0.90 & \textcolor{blue}{1.56} & \textcolor{red}{3.20} \\
    \midrule
    NAIMA (Ours) & \textcolor{blue}{1.10} & \textcolor{blue}{2.39} & \textcolor{blue}{4.61} & \textcolor{red}{1.07} & \textcolor{red}{1.61} & \textcolor{red}{2.47} & \textcolor{red}{1.03} & \textcolor{blue}{1.62} & \textcolor{red}{2.75} & 0.84 & \textcolor{red}{1.44} & \textcolor{blue}{3.34} \\
    \bottomrule
    \end{tabular}
    \caption{Results of comparison of existing techniques with NAIMA, across 4 largely used evaluation datasets, and 3 different scaling factors. All baseline results are reported from prior works, where models are trained on the NYU\_v2 dataset. 
    Total examples for each dataset are as follows: NYU\_v2: 449, RGBDD: 405, Middlebury: 30, LU: 6. Best results in \textcolor{red}{red}, second-best in \textcolor{blue}{blue}. NYU\_v2 is the in-distribution setting; the remaining datasets are out-of-distribution dataset evaluations. NAIMA remains competitive in-distribution and obtains the most best or second-best results in the cross-dataset setting, particularly at larger scaling factors.}
    \label{tab:main_res}
    
\end{table*}

We additionally evaluate the performance of our proposed NAIMA on the relatively newer TOFDSR dataset, which has not been used for comparative analysis in previous works. For a fair and transparent comparison, we selected SGNet and DCTNet, both reporting competitive performance on standard benchmarks and providing publicly available pretrained weights. The results of the comparison are presented in Table~\ref {tab:tofdsr_res}.

\begin{table}[!htb]
\centering
    \begin{tabular}{l|ccc}
    \toprule
    \textbf{Methods} & \textbf{x4} & \textbf{x8} & \textbf{x16} \\
    \midrule
    DCTNet 2022 & \textcolor{blue}{2.6} & 4.33 & 6.34 \\
    SGNet 2024 & 2.11 & \textcolor{blue}{3.27} & \textcolor{blue}{4.73} \\
    NAIMA (Ours) & \textcolor{red}{2.09} & \textcolor{red}{3.18} & \textcolor{red}{4.72} \\
    \bottomrule
    \end{tabular}
    \caption{Comparison on the TOFDSR dataset (2024), which contains 560 pairs for validation. For fair evaluation, only methods with publicly available pretrained weights are included. NAIMA outperforms the existing architectures for all three scaling factors.}
    \label{tab:tofdsr_res}
    
\end{table}

These results validate our hypothesis that incorporating semantic priors effectively enhances reconstruction quality and reduces errors introduced by noisy RGB guidance.

\subsection{Qualitative Results}
\begin{figure*}[!htb]
  \centering
  \includegraphics[width=0.93\linewidth]{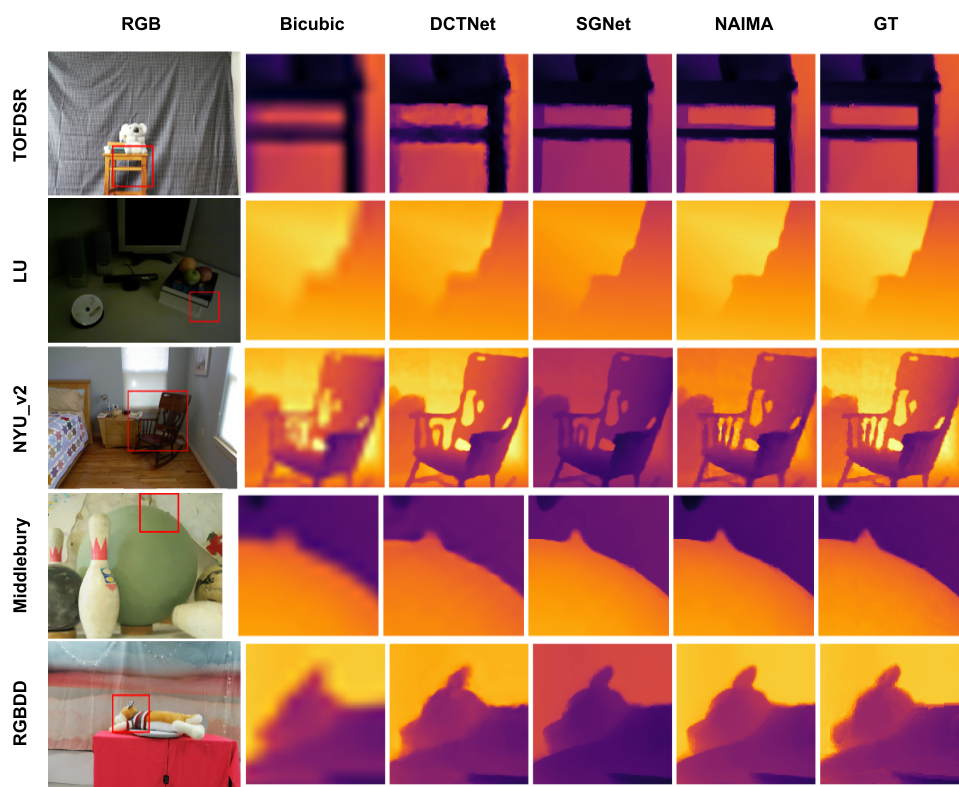}
  \caption{Comparison of depth maps at 8$\times$ scaling factor across multiple evaluation datasets. Bicubic denotes the bicubic-upsampled low-resolution input, while GT represents the ground-truth depth map. The bounding box denotes the corresponding area in the RGB image. It can be observed that NAIMA not only improves depth boundary reconstruction but also adheres to structural details while reconstructing the depth maps.}
    \label{fig:quality}
\end{figure*}

Figure~\ref{fig:quality} presents a qualitative comparison of our proposed method with existing benchmark techniques across multiple datasets. It can be observed that NAIMA recovered finer structural details while being robust to RGB feature noise. Particularly, it performs better near depth boundaries, producing sharper edges and well-defined corner structures, solidifying the use of semantic priors in mitigating the impact of RGB noise and cross-modal misalignment, which often lead to blurred depth boundaries in competing approaches.

\begin{figure*}[!htb]
  \centering
  \includegraphics[width=0.95\linewidth]{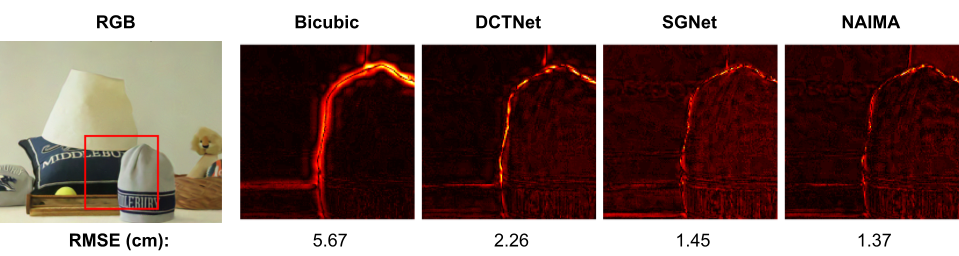}
  \caption{Comparison of error maps for a selected patch from the Middlebury dataset at 8$\times$ upscaling. The bounding box indicates the corresponding region in the RGB image.}
    \label{fig:error}
\end{figure*}

Figure~\ref{fig:error} presents the error map for the 8$\times$ scaling factor, for a selected depth boundary patch from the Middlebury dataset. The figure illustrates the error distribution, along with the RMSE reported in $cm$, where NAIMA achieves the lowest error among the techniques compared. Additional qualitative analysis can be found in the appendix.

\subsection{Intermediate Feature Analysis}
Figure~\ref{fig:feature} presents a side-by-side comparison of the RGB encoded features and their refined counterparts after processing through the GTA module at different levels of the NAIMA architecture. It can be observed that the features extracted directly from RGB contain noise that often corresponds to false depth boundaries, which are curbed by the GTA module, producing smoother and more coherent object boundaries. This refinement enables NAIMA to produce sharper depth boundaries while mitigating the adverse effects of RGB noise.

\section{Ablation}
Our ablation consists of validating the importance of semantic attention injection into depth features, the contribution of the semantic priors themselves, and verifying the effectiveness of the L1-gradient loss function. All variants are trained on NYU\_v2 under the same configuration described previously, at an 8$\times$ scaling factor to keep the study manageable, and evaluated across all five datasets.

\textbf{Attention mechanism (NAIMA$^{+}$):} To test whether attention-based injection is necessary, or whether the tokens could simply be added, we retain the semantic tokens but replace the cross-attention in Eq.~\ref{eq:residual_attention} with element-wise addition:

\begin{equation}
    D^{*}_{i} = E_{i} + F_{i}.
\end{equation}

\textbf{Semantic priors (NAIMA$^{-}$):} To quantify the contribution of the semantic priors themselves, we remove the entire semantic path, i.e., the DINOv2 tokens and the token-injection step are dropped. The intention here is to verify the gains stemming from the semantic priors.

\textbf{Gradient loss (NAIMA-L1):} To assess the gradient-aware loss, we train NAIMA with the $\mathcal{L}_{1}$ reconstruction term only, removing $\mathcal{L}_{grad}$.

Table~\ref{tab:ablation} reports the results. Removing the semantic branch degrades performance consistently across all five datasets, confirming that the pretrained tokens carry useful guidance. More revealing, however, is that replacing cross-attention with element-wise addition performs worse than removing the semantics entirely. This asymmetry is the core justification for our design stating that semantic priors are beneficial only when the depth stream can selectively query them, which is precisely what the gated cross-attention provides. Finally, removing the gradient term causes a smaller but consistent degradation, indicating its role in preserving structural detail is complementary to, and independent of, the semantic guidance.

\begin{table}[!htb]
\centering
\small
\setlength{\tabcolsep}{2.7pt}
    \begin{tabular}{lccccc}
    \toprule
    \textbf{Methods}
    & \textbf{Middlebury}
    & \textbf{LU}
    & \textbf{NYU\_v2}
    & \textbf{RGBDD} 
    & \textbf{TOFDSR} \\
    \midrule
    NAIMA & 1.62 & 1.44 & 2.39 & 1.61 & 3.18 \\
    NAIMA-L1 & 1.67 & 1.45 & 2.42 & 1.63 & 3.22 \\
    NAIMA$^{-}$ & 1.70 & 1.50 & 2.49 & 1.68 & 3.26 \\
    NAIMA$^{+}$ & 1.73 & 1.50 & 2.51 & 1.72 & 3.28 \\
    \bottomrule
    \end{tabular}
    \caption{Ablation of NAIMA components, trained on NYU\_v2 at $8\times$ and evaluated across all datasets. NAIMA$^{+}$ replaces cross-attention with element-wise addition, NAIMA$^{-}$ removes the semantic branch entirely, and NAIMA-L1 removes the gradient term from the loss. Every variant degrades relative to the full model, confirming the importance of semantic priors, the attention-based injection, and the gradient loss.}
    \label{tab:ablation}
    
\end{table}

\section{Conclusion}
We introduce a novel architecture for guided depth super-resolution (GDSR) that incorporates additional semantic priors derived from pretrained vision transformer embeddings. Unlike existing approaches, our framework leverages semantically rich token representations as an explicit modality for guided depth super-resolution, enabling more faithful utilization of RGB information. Our architecture remains competitive with state-of-the-art methods in-distribution while achieving the strongest cross-dataset generalization at larger scaling factors, without decoded priors or auxiliary objectives. Furthermore, we propose a novel Guided Token Attention (GTA) module that selectively transfers relevant semantic information across multiple levels of token abstraction into the depth representation. By progressively aligning depth features with semantic tokens of varying complexity, the model learns to perform more reliable GDSR, thereby reducing the adverse effects of RGB texture and color noise, which often lead to erroneous depth discontinuities.

{
    \small
    \bibliographystyle{ieeenat_fullname}
    \bibliography{main}
}

\appendix
\clearpage
\setcounter{page}{1}
\section*{Appendix}
\label{sec:appendix}
We present additional implementation and analytical details here, summarized as follows:
\begin{itemize}
    \item Appendix~\ref{sec:experimental}: Experimental setup and training details.
    \item Appendix~\ref{sec:quality_more}: Additional qualitative analysis.
\end{itemize}

\section{Experimental Details}
\label{sec:experimental}
\subsection{Loss and Convergence}
\begin{figure}[!htb]
  \centering
  \includegraphics[width=0.9\linewidth]{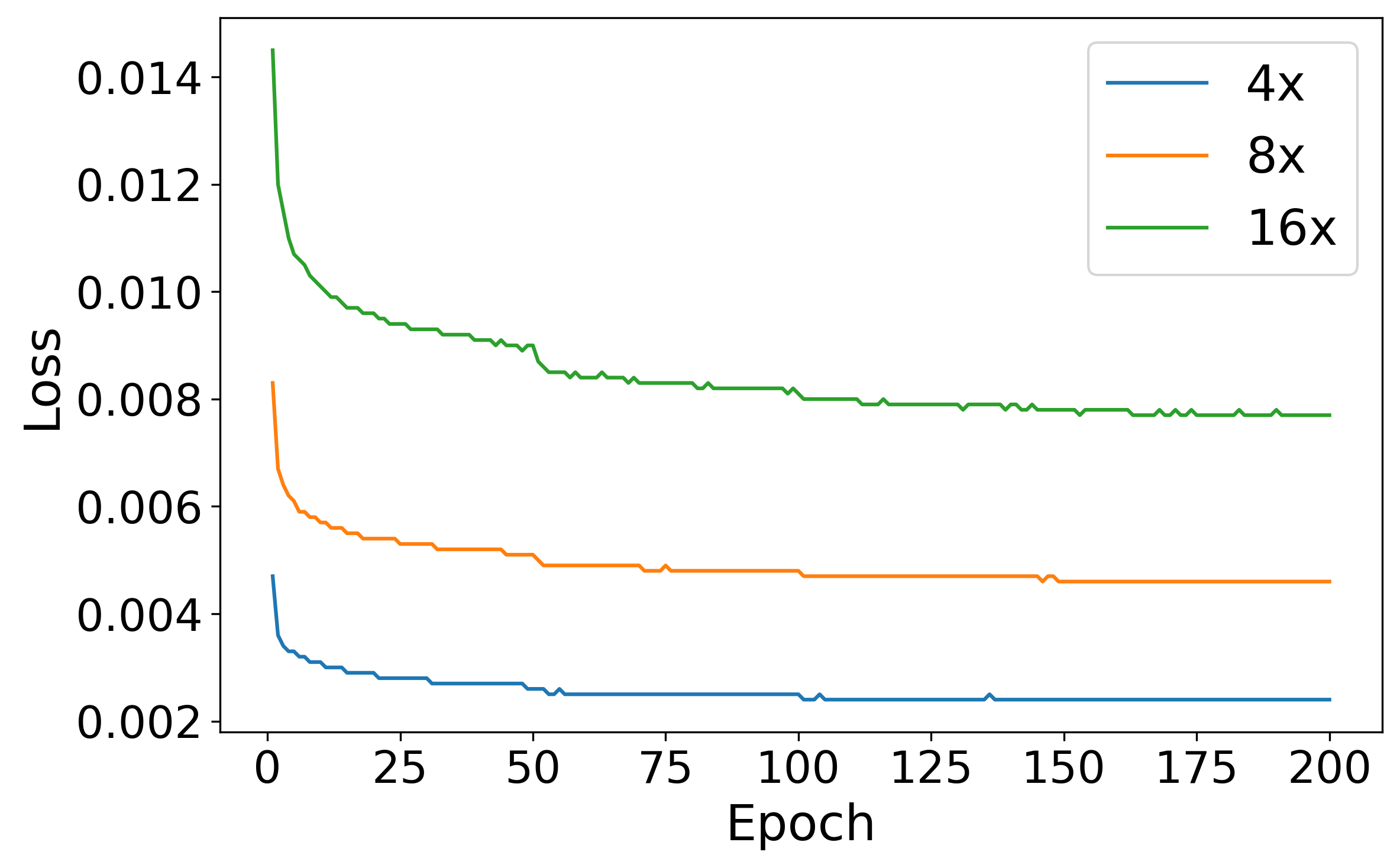}
  \caption{Loss plots for NAIMA models at scaling factors of 4x, 8x, and 16x. All three models converge well within the 200 epochs.}
    \label{fig:loss}
\end{figure}

We first present an overview of the training loss observed for the 3 versions of NAIMA. Figure~\ref{fig:loss} illustrates the loss curves for the three scaling factors 4x, 8x, and 16x. All models exhibit stable convergence over 200 training epochs.

\subsection{Semantics Encoder Settings}
As mentioned, we used DINOv2 as the semantic feature extractor for the input RGB image. Specifically, we used the ViT-S/14 variant containing 22 million parameters. The model is pretrained on a diverse corpus consisting of 142 million unlabeled images, capturing robust semantic features that benefit our downstream task of depth super-resolution. The encoder’s weights are kept frozen during training to preserve these generalized, rich semantic representations learned from the large-scale pretraining. We extracted semantic tokens from the 2nd, 5th, 8th, and 11th transformer blocks of DINO, providing a balanced representation of low- and high-level semantic information.

\subsection{Model Complexity Details}
Table~\ref{tab:complexity} presents the model complexity across different scaling factors. We report the total number of parameters, along with a breakdown of learnable and non-learnable parameters, and FLOPs for each scale. The majority of the non-learnable parameters are from the semantic encoder.

\begin{table}[!htb]
\centering
\small
\setlength{\tabcolsep}{2.7pt}
    \begin{tabular}{lcccc}
    \toprule
    \textbf{Scale} & \textbf{Params (M)} & \textbf{Train. (M)} & \textbf{Non-Train. (M)} & \textbf{FLOPs (T)} \\
    \midrule
    4$\times$  & 61.459  & 40.968  & 21.876  & 2.722 \\
    8$\times$  & 80.260  & 59.769  & 21.924  & 5.009 \\
    16$\times$ & 140.422 & 119.931 & 21.924  & 11.374 \\
    \bottomrule
    \end{tabular}
    
\caption{Model complexity of NAIMA across different scaling factors. Parameters are reported in millions (M), and FLOPs in tera floating point operations (T).}
\label{tab:complexity}
\end{table}

\subsection{Additional Implementation Choices}
Next, we discuss additional configurations, including patch size selection and training protocols.

\textbf{Patch Size:}
We use a patch size of 420 for 4$\times$ scaling factor, and 448 for 8$\times$ and 16$\times$ scaling factors. These values are selected to ensure compatibility with the DINOv2 semantics encoder, which requires the input image size to be divisible by 14 for its patch tokenization. Additionally, a smaller patch size was chosen for the 4$\times$ scaling factor due to GPU memory limitations.

\textbf{Training Protocols:} We employ a step-based learning rate decay scheduler, using 0.3 as the decay factor, reducing the learning rate after every 50 epochs. All RGB images are standardized using ImageNet normalization statistics. For reproducibility, the complete data processing pipeline and training code are provided in the code repository 
\footnote{\href{https://github.com/tayyabnasir22/NAIMA-GDSR}{NAIMA} and \href{https://github.com/tayyabnasir22/GDSR-Data-Preperation}{data generator code}.}

\textbf{Inference Time Padding:}
As aforementioned, our model required input images having a size divisible by 14 for the DINOv2 encoder. During inference, the input images whose spatial dimensions are not divisible by 14 are padded with additional zeros at the bottom right corner before being processed by the model. After the inference, the predicted depth map is cropped back to the original resolution, prior to being evaluated using RMSE. This step is necessary to satisfy the architectural constraints of DINOv2, although it may introduce a minor bias in the reported RMSE due to padding artifacts, slightly adding to the reported error.

\section{Qualitative Comparisons}
\label{sec:quality_more}
\begin{figure*}[!htb]
  \centering
  \includegraphics[width=0.85\linewidth]{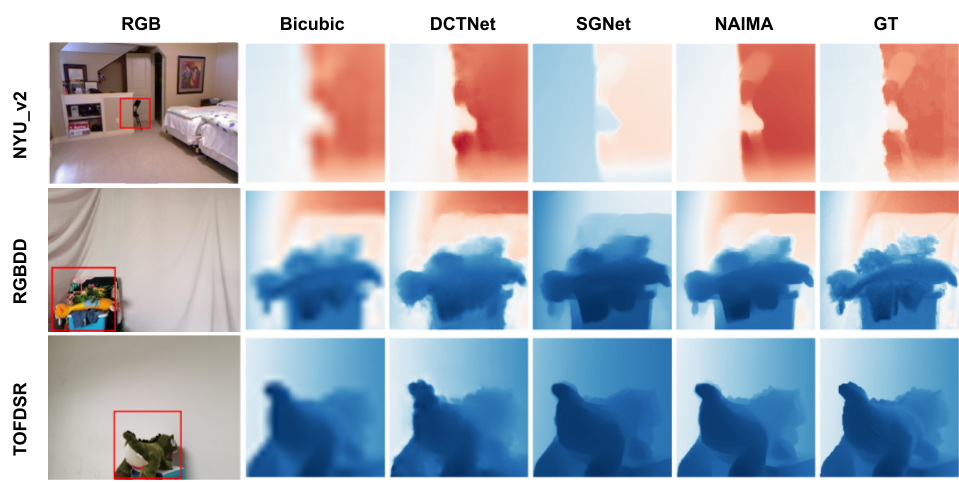}
  \caption{Comparison of depth maps at 16$\times$ scaling factor across multiple evaluation datasets. Bicubic denotes the bicubic-upsampled low-resolution input, while GT represents the ground-truth depth map. The bounding box denotes the corresponding area in the RGB image. For the 16$\times$ scale, where the low-resolution depth input contains severely limited structural information, and most cues must be inferred from the RGB guidance, NAIMA successfully reconstructs sharp depth discontinuities and recovers fine structural details. Notably, it accurately captures the chair structure in the first example, preserves the object's edge boundaries in the second, and finely reconstructs the fins on the toy in the third.}
    \label{fig:16x_quality}
\end{figure*}

\begin{figure*}[!htb]
  \centering
  \includegraphics[width=0.85\linewidth]{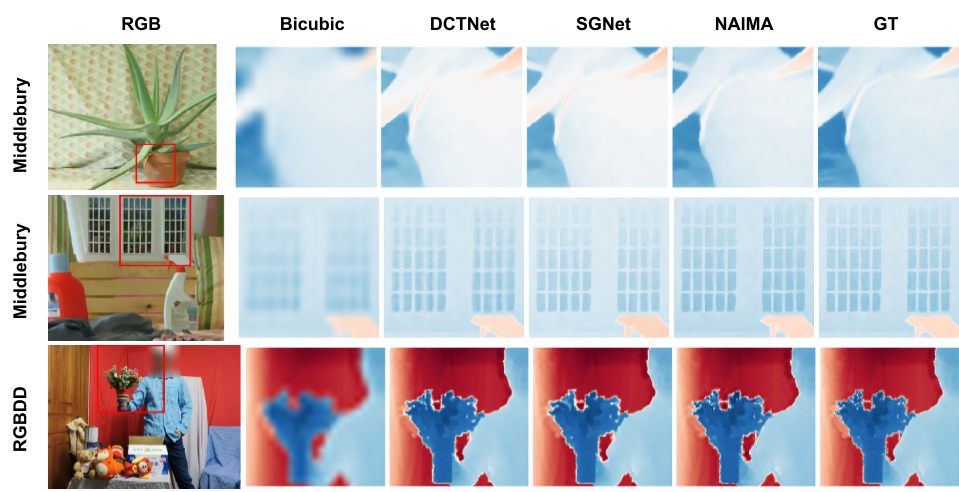}
  \caption{Comparison of depth maps at 4$\times$ scaling factor across multiple evaluation datasets. Bicubic denotes the bicubic-upsampled low-resolution input, while GT represents the ground-truth depth map. The bounding box denotes the corresponding area in the RGB image. NAIMA provides more accurate depth estimation for fine structures, such as the thin plant vine in the first example. In the second example, it better preserves depth discontinuity in the window’s structure compared to competing methods. In the third example, NAIMA reconstructs sharper and more precise depth boundaries for the flower pattern, closely matching the ground truth.}
    \label{fig:4x_quality}
\end{figure*}

Finally, we present a qualitative comparison of multiple scaling factors. While the main paper focuses only on results at 8$\times$, Figures~\ref{fig:16x_quality} and ~\ref{fig:4x_quality} provide additional comparisons for 16$\times$ and 4$\times$ scaling factors, respectively. It is evident from both the figures that NAIMA consistently achieves sharper depth discontinuities and more accurate reconstruction of fine-grained structures. 

\end{document}